# A Machine Vision Approach to Preliminary Skin Lesion Assessments

Quinn McGill, Ali Khreis, Ro'Yah Radaideh
University of Ottawa

**Abstract** –Early detection of malignant skin lesions is critical to improving patient outcomes for aggressive, metastatic skin cancers. This paper first details a machine vision system which classifies skin lesion images from a subset of the HAM 10,000 dataset into benign and malignant classes based on the clinically established ABCD rule of dermoscopy. By analyzing a lesion's Asymmetry, Borders, Color, and Dermoscopic Structures, the system acts as an efficient stand-in for a preliminary dermatological assessment. The scores for each category are combined into a Total Dermatology Score (TDS) to generate the malignancy assessment. This approach was then compared with solutions which leverage machine learning. The handcrafted ABCD features were fed to classification models like Logistic Regression and Random Forest to generate malignancy labels. EfficientNet-B0, a large-scale image classification model was fine-tuned on the 1,000-image subset. Finally, a Convolutional Neural Network was fully trained to classify lesions as benign or malignant. The results of the study suggest that while the traditional, TDS-based classification system offers more clinically interpretable results, it is limited in its final classification accuracy. The transfer learning approach had very poor performance as a result of domain shift. Despite the small dataset size, the fully trained CNN offered the best performance across all metrics.

## I. INTRODUCTION

Skin cancer is one of the most common forms of cancer, with 9,500 people diagnosed every day in the United States [1]. The deadliest form of skin cancer is malignant melanoma, which can develop from regular-looking moles and appear on any area of the body, including those which are not regularly exposed to the sun [2]. However, when detected early, the 5-year survival rate of melanoma is 99 percent [1]. The first step in patient care is a preliminary dermatology assessment to determine a lesion's level of suspicion. Dermoscopy is a non-invasive medical technique which uses a dermatoscope to produce magnified images of skin to assess structures that are not visible to the naked eye. Dermatologists rely on dermoscopic images during a preliminary assessment to determine whether a lesion is malignant or benign. Many designs have been proposed for a Computer-Aided Diagnosis (CAD) system which can expedite preliminary malignancy assessment, enabling earlier treatment, thereby improving patient outcomes. Many rely on traditional image processing techniques such as image segmentation and histogram analysis, while more modern solutions integrate machine learning models to improve classification accuracy.

## II. LITERATURE REVIEW

Saba [3] compared the efficacy of handcrafted (traditional image processing) and non-handcrafted (deep learning-based) features for skin cancer diagnosis. Images from large dermatology datasets like ISIC were preprocessed using noise removal techniques, including median and Gaussian filtering to remove hairs and blood vessels which can impact lesion segmentation. Traditional image processing techniques were used to generate handcrafted features to match clinical tools used during a dermatological assessment, including the 7-point checklist and ABCD characteristics.

Mahmoud and Soliman [4] developed a Computer-Aided Diagnosis system for skin cancer which used artificial intelligence for improved lesion segmentation and malignancy classification. Their findings suggest that an Adaptive Snake algorithm can produce extremely precise lesion boundaries, and Artificial Neural Networks achieve high classification accuracy.

Liu et al. [5] provided a review of modern computer vision techniques and deep learning models which are used in CAD systems for early detection of melanoma. Generative Adversarial Networks can be leveraged for data augmentation in medical datasets, while models like You Only Look Once (YOLO) and Mask R-CNN are great for lesion segmentation. Convolutional Neural Networks (CNNs) stand out due their ability to automatically extract features from dermoscopic images and classify lesions as benign or malignant.

## III. METHODOLOGY

### A. Dataset Preparation and Preprocessing

The "Human Against Machine with 10,000 training images" or HAM10000 dataset is a collection of over 10,000 dermoscopic images from different populations and is widely used in the design of CAD systems for skin cancer. Seven diagnoses are represented: Actinic keratoses and intraepithelial carcinoma / Bowen's disease, basal cell carcinoma, benign keratosis-like lesions, dermatofibroma, melanoma, melanocytic nevi, and vascular lesions [6]. A smaller, more manageable subset (1,000 images) of the dataset was created, consisting of 50% malignant and 50% benign lesions. To reduce the effect of hairs, blood vessels and blood on lesion segmentation, three noise removal filtering techniques were assessed: flat average (3x3 kernel), Gaussian (3x3 kernel and sigma value 1.0), and median (3x3). This generated three versions of the 1,000-image subset with a subtle blurring effect. While hair and other noise were blurred in the filtered images, the lesion boundaries were largely unaffected by pre-processing.

B. Machine Vision Approach

The machine vision component of this system is designed to mimic the preliminary assessment of a dermatologist by quantitatively analyzing the four key characteristics defined by the ABCD rule: Asymmetry, Border irregularity, Color diversity, and Dermoscopic structures. These four features are extracted from the lesion image using a custom computer vision pipeline. The individual feature scores are then combined into a single diagnostic metric, the Total Dermoscopy Score (TDS), to provide a final malignancy assessment. This structured, rule-based feature extraction provides the input data for the traditional machine learning classifiers detailed in the next section.

The initial and most critical step is lesion segmentation, which isolates the cancerous area from the surrounding healthy skin. This process uses Otsu's thresholding applied to the grayscale, Gaussian-blurred image, based on the assumption that the lesion is significantly darker than the background, which is the case for the subset of images used. Following thresholding, standard morphological operations are applied to smooth the mask, fill small internal holes, and eliminate spurious noise. Finally, the largest detected contour is isolated and filled to ensure that only the primary lesion is retained, producing a clean binary mask for subsequent feature extraction. An example of lesion segmentation is shown in Figure 1.

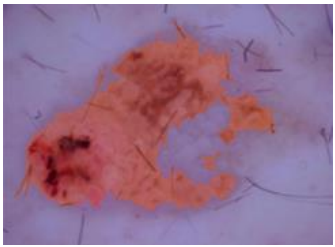

Fig. 1. Segmentation of lesion from surrounding health skin

The Asymmetry score (A) quantifies the degree to which the lesion deviates from reflectional symmetry. This is achieved by first identifying the primary axis of the lesion using Principal Component Analysis (PCA) on the coordinates of all pixels within the segmented mask. PCA efficiently finds the axis of least asymmetry, also known as the major axis, which is then used to orient the lesion horizontally via rotation.

The rotated lesion is split into two halves along its major and minor axes, creating four quadrants. The asymmetry is measured along each axis by calculating the Intersection over Union (IoU) between one half and the reflection of the other half. An IoU of 1.0 indicates perfect symmetry (no difference), while a lower value signifies a greater difference. The asymmetry for each axis is calculated as 1 - IoU. The final A-score ranges from 0 to 2, where a score of 1 is assigned if the asymmetry (difference) exceeds a threshold (t=0.15) on one axis, and a score of 2 is given if the threshold is surpassed on both axes.

Border Irregularity (B) is assessed by checking for abrupt transitions in pixel intensity at the boundary of the lesion, which are characteristic of aggressive growth. The algorithm examines the lesion border along eight equally spaced radial segments originating from the lesion's center of mass.

For each segment, the boundary is identified as the last pixel inside the lesion mask along the radial line. The gradient is calculated by comparing the average intensity of a small patch of pixels immediately inside the border with a patch immediately outside the border. An abrupt border segment is flagged if this intensity gradient exceeds a predefined threshold (set to 10 in the code). The B-score is the total count of these abrupt segments, with a maximum possible value of 8.

The Color Diversity score (C) reflects the number of distinct colors present within the lesion, a strong indicator of malignancy. To improve perceptual color difference detection, the pixels within the lesion mask are first converted from the BGR (Blue, Green, Red) color space to the perceptually uniform LAB color space.

K-Means clustering is then applied to group the lesion's pixels into up to five color clusters. To focus only on diagnostically significant colors, only clusters that represent at least 5 of the total lesion area are considered. An example of color clustering is shown in Figure 2. Finally, an inter-cluster distance check is performed to merge any of these significant clusters whose centers are too close (distance < 10 in LAB space), ensuring only truly distinct colors contribute to the score. The C-score is the count of these unique, significant color clusters, capped at a maximum of 6.

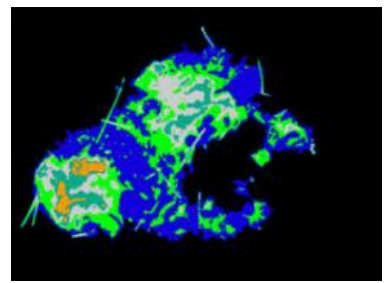

Fig. 2. Segmentation of lesion from surrounding health skin

The Dermoscopic Structures score (D) identifies the presence of four key patterns, which sum up to a maximum score of 4:

1. Structureless Areas: Detected if the median local pixel variance inside the mask is below a low threshold of 20, indicating a homogeneous region.
2. Dots/Globules: Identified using the Laplacian of Gaussian (LoG) blob detection method. The structure

is counted if 3 or more circular 'blobs' are found within the lesion mask.
3. Pigment Network: Detected by applying a local threshold to the image to isolate dark pigment areas, followed by skeletonization to reduce lines to 1-pixel thickness. A network is confirmed if the number of branch points (pixels connected to more than three neighbors on the skeleton) exceeds a threshold 20.
4. Streaks/Pseudopods: Detected as linear features using the Probabilistic Hough Line Transform on Canny-detected edges. The structure is counted if a sufficient number of line segments (more than 5) are identified.

The D-score is the simple sum of the binary presence (1 or 0) of these four structures.

The extracted ABCD features are combined using the original clinically derived formula to calculate the Total Dermoscopy Score (TDS):

$$TDS = 1.3 * A + 0.1 * B + 0.5 * C + 0.5 * D \quad (1)$$

The weights emphasize Asymmetry (1.3) and give equal importance to Color and Differential Structures (0.5), while the Border score carries the least weight (0.1). The final malignancy assessment is derived from the calculated TDS, classifying the lesion into one of three categories:

- Benign: TDS < 4.75
- Suspicious: 4.75 ≤ TDS ≤ 5.45
- Malignant: TDS > 5.45

The resulting four feature scores (A, B, C, D) and the final TDS are the primary feature set used for evaluation in the traditional machine learning approach.

The machine vision component implemented an automated, rule-based system for quantifying the diagnostic characteristics defined by the ABCD dermoscopy rule, resulting in a structured, five-dimensional feature vector (A, B, C, D, and TDS). This approach provides a transparent and clinically grounded output, confirming that the median-filtered preprocessing stream was the most suitable choice, as it best preserved the necessary edge information for accurate geometric (A-score) and gradient (B-score) analysis. Nevertheless, the subsequent evaluation using traditional ML classifiers demonstrated that this handcrafted feature set achieved only modest results (best recall of 61.0%). This performance indicates a significant bottleneck in the pipeline: the reduction of complex, high-resolution visual morphology into only five numerical values fundamentally limits the information available to the downstream classifiers, failing to capture the full spectrum of subtle patterns required for reliable malignancy detection.

C. Machine Learning Approach

The machine learning component of this project performs binary classification of skin lesions using two approaches: traditional classifiers on handcrafted ABCD features, and convolutional neural networks trained directly on the images.

Three classifiers were evaluated for use on ABCD features: Logistic Regression, Random Forest, and SVM. The input consisted of five features: the four ABCD scores (Asymmetry 0-2, Border 0-8, Color 1-6, Structures 0-5) and the Total Dermoscopy Score. Features were z-score normalized, and class imbalance was addressed using inverse-frequency class weights. An 80/20 stratified split was used, with each classifier evaluated across all three preprocessing streams (median, Gaussian, flat average).

The first dermoscopic image classification approach used EfficientNet-B0 pretrained on ImageNet with a freeze-then-finetune strategy: the backbone was frozen for 10 epochs while training the classification head (Global Average Pooling, Dropout Regularization, then Dense Layer), then unfrozen for 30 epochs at reduced learning rate (1e-5). Data augmentation (rotation, flips, shifts, zoom, brightness) and class weights were applied to reduce overfitting and address class imbalance.

As a result of poor performance with the fine-tuned model, a custom CNN was developed for the limited 1,000-image dataset. The architecture consists of three convolutional blocks (16→32→64 filters) with 3×3 kernels, ReLU activations, and 2×2 max pooling, reducing spatial dimensions from 224×224 to 28×28. The output is flattened and passed through fully connected layers (64 units → 1 output). The model was implemented in PyTorch with Binary Cross Entropy (BCE) loss and Adam optimizer (lr=1e-3) for 10 epochs.

IV. RESULTS

The performance of all classification approaches was evaluated using five standard metrics: accuracy (overall correctness), precision (positive predictive value), recall (sensitivity to malignant cases), F1-score (harmonic mean of precision and recall), and ROC-AUC (area under the receiver operating characteristic curve). For clinical screening applications, recall is particularly critical as false negatives (missed malignancies) carry severe consequences. Results are summarized in Tables 1 and 2.

Table 1 presents the classification performance of traditional machine learning models trained on ABCD features across the three preprocessing streams.

Table 1: Traditional ML Performance on ABCD Features

| Stream | Model | Acc | Prec | Rec | F1 | AUC |
|---|---|---|---|---|---|---|
| Med | SVM | 0.590 | 0.587 | 0.610 | 0.598 | 0.619 |
| Med | LR | 0.595 | 0.590 | 0.620 | 0.605 | 0.590 |
| Med | RF | 0.595 | 0.592 | 0.610 | 0.601 | 0.599 |
| Gaus | SVM | 0.585 | 0.583 | 0.600 | 0.591 | 0.593 |
| Flat | RF | 0.585 | 0.571 | 0.680 | 0.621 | 0.598 |

The median-filtered stream consistently outperformed Gaussian and flat average across most metrics, suggesting median filtering best preserves edge information critical for ABCD feature extraction. SVM achieved the highest ROC-AUC (0.619) while Logistic Regression achieved the highest accuracy (59.5%). Overall, traditional classifiers achieved modest performance (54.5-59.5% accuracy), limited by the five-dimensional feature space.

Transfer learning with EfficientNet-B0 failed critically, despite 80% precision; it achieved only 4% recall, missing 96% of malignant cases by defaulting to the majority class. The pretrained ImageNet features did not transfer effectively, and the limited dataset was insufficient for adapting 4+ million parameters.

In contrast, the custom CNN trained from scratch achieved substantially improved performance across all metrics. Table 2 summarizes the results.

Table 2: CNN Classification Performance

| Stream | Model | Acc | Prec | Rec | F1 | AUC |
|---|---|---|---|---|---|---|
| Med | Transfer | 0.515 | 0.800 | 0.040 | 0.076 | 0.615 |
| Med | Custom CNN | 0.785 | 0.756 | 0.865 | 0.807 | 0.801 |
| Flat | Custom CNN | 0.775 | 0.727 | 0.842 | 0.780 | 0.833 |
| Gaus | Custom CNN | 0.750 | 0.767 | 0.704 | 0.734 | 0.803 |

The median-filtered stream with the custom CNN achieved the highest accuracy (78.5%) and recall (86.5%), correctly identifying the vast majority of malignant cases. The flat average stream achieved the highest ROC-AUC (0.833), indicating strong overall discriminative ability.

The custom CNN outperformed the best traditional ML classifier by 19 percentage points in accuracy (78.5% vs. 59.5%) and 25 percentage points in recall (86.5% vs. 61.0%). This improvement demonstrates that direct pixel-level learning captures diagnostic patterns beyond what the handcrafted ABCD features encode. The success of the custom CNN versus the failure of transfer learning highlights that for small, domain-specific datasets, a lightweight architecture trained from scratch can outperform sophisticated pretrained models.

## V. CONCLUSION

This study presented a comprehensive system for preliminary skin lesion assessment, combining the clinically established ABCD dermoscopy rule with machine learning classification. The system was evaluated on a 1,000-image subset of the HAM10000 dataset across three preprocessing streams: median, Gaussian, and flat average filtering.

The machine vision component successfully implemented an automated, rule-based pipeline for extracting the four ABCD features; Asymmetry, Border irregularity, Color diversity, and Dermoscopic structures; and computing the Total Dermoscopy Score (TDS). This approach provides transparent, clinically interpretable outputs that mirror dermatological practice. The median-filtered preprocessing stream proved most effective, as it best preserved edge information critical for accurate geometric and gradient analysis.

However, the evaluation revealed a fundamental limitation of the handcrafted feature approach. Traditional ML classifiers trained on the five-dimensional ABCD feature vector achieved only modest performance, with the best model (Logistic Regression on median-filtered images) reaching 59.5% accuracy and 62.0% recall. This performance bottleneck stems from the reduction of complex, high-resolution visual morphology into only five numerical values, which fails to capture the full spectrum of subtle patterns required for reliable malignancy detection.

The deep learning investigation yielded two contrasting outcomes. Transfer learning with EfficientNet-B0 pretrained on ImageNet failed critically, achieving only 4% recall despite 80% precision. This demonstrates that pretrained features from natural images do not transfer effectively to the specialized dermoscopic domain, particularly with limited training data. In contrast, a custom three-layer CNN trained from scratch achieved 78.5% accuracy and 86.5% recall on median-filtered images; a 19-point accuracy improvement and 25-point recall improvement over the best traditional classifier.

These findings support three key conclusions. Firstly, preprocessing methodology significantly impacts classification performance, with median filtering consistently outperforming other methods. Moreover, while ABCD-based feature extraction provides valuable clinical interpretability, its dimensional reduction creates an information bottleneck that limits classification accuracy. Finally, for small, domain-specific medical imaging datasets, purpose-built lightweight architectures can substantially outperform both handcrafted features and transfer learning from unrelated domains.

## VI. FUTURE WORK

Several observations about future improvements emerge from this study. Firstly, increasing the dataset size to the entirety of the HAM10000 dataset (10,015 images) or combining multiple benchmark dermatological datasets (ISIC Archive, PH2) would significantly improve the generalizability and overall performance of machine learning-based solutions.

Going forward, hybrid ensemble architectures combining ABCD features with CNN pattern recognition could combine interpretability and learning power simultaneously.

Additionally, domain-specific transfer learning using models pretrained on medical imaging datasets or self-supervised techniques (SimCLR, MoCo) on unlabeled dermoscopic images, may provide more relevant feature representations. Explainability mechanisms such as Grad-CAM are essential for clinical adoption, allowing dermatologists to verify the model focuses on clinically relevant regions.

Finally, multi-class classification extending beyond binary labels to the seven HAM10000 categories would better align with clinical workflows. Finally, clinical validation studies comparing system recommendations against expert diagnoses are necessary before deployment.

## REFERENCES

[1] "Skin cancer facts & statistics," The Skin Cancer Foundation, https://www.skincancer.org/skin-cancer-information/skin-cancer-facts/ (accessed Dec. 5, 2025).

[2] "Skin cancer," Canadian Skin Cancer Foundation, https://www.canadianskincancerfoundation.com/skin-cancer/ (accessed Dec. 5, 2025).

[3] T. Saba, "Computer vision for microscopic skin cancer diagnosis using handcrafted and non-handcrafted features," *Microscopy Research and Technique*, vol. 84, no. 6, pp. 1272–1283, Jan. 2021. doi:10.1002/jemt.23686

[4] N. M. Mahmoud and A. M. Soliman, "Early Automated Detection System for skin cancer diagnosis using artificial intelligent techniques," *Scientific Reports*, vol. 14, no. 1, Apr. 2024. doi:10.1038/s41598-024-59783-0

[5] Y. Liu et al., "Advances in computer vision and deep learning-facilitated early detection of melanoma," *Briefings in Functional Genomics*, vol. 24, Mar. 2025. doi:10.1093/bfgp/elaf002

[6] "The HAM10000 dataset, a large collection of multi-source dermatoscopic images of common pigmented skin lesions - vidir dataverse," *Harvard Dataverse*, https://dataverse.harvard.edu/dataset.xhtml?persistentId=doi:10.7910/DVN/DBW86T (accessed Dec. 5, 2025).